\documentclass[paper]{jhep3} 

\JHEPspecialurl{http://jhep.sissa.it/JOURNAL/JHEP3.tar.gz}

\usepackage{epsfig,multicol}
\usepackage{cite}

\usepackage{amsmath}
\usepackage{amssymb}
\preprint{
TIFR/TH/07-33\\
DIAS-STP-07-20\\
KEK-TH-1104\\
arXiv:0712.0646}

\title{The instability of intersecting fuzzy spheres}
\author{Takehiro Azuma${}^{a,b}$, Subrata Bal${}^{c}$ and 
Jun Nishimura${}^{a,d}$ \\
\llap{$^a$}Institute of Particle and Nuclear Studies, \\
High Energy Accelerator Research Organization (KEK),\\
1-1 Oho, Tsukuba 305-0801, Japan  \\
\llap{$^b$}Department of Theoretical Physics, Tata Institute of Fundamental Research (TIFR), \\
Homi Bhabha Road, Mumbai, 400-005, India \\
\llap{$^c$}School of Theoretical Physics, Dublin Institute for Advanced Studies (DIAS),\\
10 Burlington Road, Dublin 8, Ireland \\
\llap{$^d$}Department of Particle and Nuclear Physics, \\
Graduate University for Advanced Studies (SOKENDAI),\\
1-1 Oho, Tsukuba 305-0801, Japan \\
\email{azuma@theory.tifr.res.in, 
sbal@stp.dias.ie, 
jnishi@post.kek.jp }
}

\abstract{We discuss the classical and quantum
stability of general configurations representing 
many fuzzy spheres
in dimensionally reduced Yang-Mills-Chern-Simons models
with and without supersymmetry.
By performing one-loop perturbative calculations around
such configurations, we find that 
intersecting fuzzy spheres are classically unstable in
the class of models studied in this paper.
We also discuss the large-$N$ limit of 
the one-loop effective action 
as a function of the distance of fuzzy spheres.
This shows, in particular, that concentric fuzzy spheres
with different radii, which
are identified with the 't Hooft-Polyakov monopoles,
are perturbatively stable
in the bosonic model and in the $D=10$ supersymmetric model.
}


\newcommand{\bel}{\begin{equation}\label}
\newcommand{\f}{\frac}

\newcommand {\beq}{\begin{equation}}
\newcommand {\eeq}{\end{equation}}
\newcommand {\beqa}{\begin{eqnarray}}
\newcommand {\eeqa}{\end{eqnarray}}
\newcommand {\bc}{\begin{center}}
\newcommand {\ec}{\end{center}}
\newcommand {\tr}{{\rm tr\,}}

\newcommand {\ee}{\mbox{e}}

\def\a{\alpha}

\def\l{\lambda}

\def\e{\epsilon}

\def\dag{\dagger}

\def\vs5{\vspace*{5mm}}
\def\vs1{\vspace*{1cm}}
\def\vs2{\vspace*{2cm}}
\def\hs5{\vspace*{5mm}}
\def\hs1{\hspace*{1cm}}
\def\hs2{\hspace*{2cm}}
\def\vs50{\vspace*{50mm}}
\def\vs20{\vspace*{20mm}}

\def\tr{\hbox{tr}}

\begin{document}

\section{Introduction}
Fuzzy spheres \cite{Madore} are simple compact noncommutative manifolds 
and have been studied extensively from various motivations. 
First it is expected that
the noncommutative geometry 
provides a crucial link to string theory and 
quantum gravity.
Indeed the Yang-Mills theory on noncommutative geometry
is shown to emerge from a certain low-energy limit of 
string theory \cite{Seiberg:1999vs}.
There is also an independent observation that the space-time 
uncertainty relation, which is naturally realized by noncommutative
geometry, can be derived from some general assumptions
on the underlying theory of quantum gravity \cite{gravity}.
One may also use fuzzy sphere
as a regularization scheme alternative to 
the lattice regularization \cite{Grosse:1995ar}.
Unlike the lattice, fuzzy sphere preserves the continuous symmetries 
of the space-time considered, and
the well-known problem of chiral symmetry 
and supersymmetry in lattice theories may become easier to overcome.

As expected from 
the Myers effect \cite{Myers:1999ps} in string theory,
fuzzy spheres appear as classical solutions \cite{alekseev}
in matrix models with a Chern-Simons term\footnote{Such models 
appear also in the context of superstring theory in the 
so-called pp-wave background \cite{pp-wave}.}.
The perturbative properties of the fuzzy spheres in matrix models 
have been studied in refs.\ \cite{0101102,0108002,0207115,perturbative}.
One can actually use matrix models to define a regularized 
field theory on a fuzzy sphere \cite{0101102,0207115}. 
Such an approach has been successful
in the case of noncommutative torus \cite{AMNS}, 
where nonperturbative studies have produced 
various important results \cite{simNC}. 
These matrix models belong to the class of
so-called large-$N$ reduced models,
which are believed to provide a constructive definition of 
superstring and M theories.
For instance, the IIB matrix model \cite{9612115},
which can be obtained by dimensional reduction of 
10d 
${\cal N}=1$ super Yang-Mills theory, is proposed as 
a constructive definition of type IIB superstring theory.
In this model the space-time is represented by the eigenvalues 
of bosonic matrices, and hence treated as a dynamical object.
The dynamical generation of 4d
space-time 
has been discussed in refs.\ \cite{4dspace}.

In ref.\ \cite{0401038}
we performed the first non-perturbative study 
of the dimensionally reduced 
Yang-Mills-Chern-Simons (YMCS) model, 
which incorporates the fuzzy sphere as a classical solution. 
When the coefficient of the Chern-Simons term ($\alpha$)
is large, the fuzzy sphere appears as the true vacuum.
However, as we decrease $\alpha$, the fuzzy sphere becomes
unstable at some critical point, and the system
undergoes a first-order phase transition.
At small $\alpha$, the large-$N$ behavior of the model becomes
qualitatively the same as in the pure Yang-Mills model ($\alpha = 0$).
This work has triggered extensive studies of 
fuzzy spheres based on Monte Carlo simulation \cite{fuzzy-sim,0506062}.
See refs.\ \cite{extended-fuzzy}
for studies on other fuzzy manifolds.

In fact the dimensionally reduced YMCS model
also has classical solutions which describe many fuzzy spheres.
These configurations include concentric fuzzy 
spheres
as well as intersecting ones.
Concentric fuzzy spheres appear in the context 
of the dynamical generation of non-trivial gauge groups 
in matrix models \cite{dyn-gauge}.
They are also used to construct 
${\bf R} \times \rm{S}^3$ geometry \cite{S3},
which is important in the context of the AdS/CFT correspondence.
Intersecting fuzzy spheres, on the other hand,
is interesting from the viewpoint of 
the brane world scenario \cite{Blumenhagen:2005mu}.

The aim of this paper is to study the classical and quantum
stability of such configurations.
At the one-loop level, it suffices to consider the interaction
between two fuzzy spheres which have different 
radii and centers in general.
The one-loop effective action around the two-fuzzy-sphere 
configuration has been calculated previously in ref.\ \cite{0108002},
and the asymptotic behaviors for large separation
and for small separation have been discussed.
We extend this study in the following directions.
Firstly we discuss the classical instability of intersecting
fuzzy spheres, which appears for intermediate separation
and hence was completely overlooked in ref.\ \cite{0108002}.
Secondly we discuss the quantum (in)stability of separate
fuzzy spheres. Here our results include the results in
ref.\ \cite{0108002}, but we further take the large-$N$ limit
and confirm that the conclusion remains unaltered.
Thirdly we extend our analysis to higher-dimensional supersymmetric 
models, and find in particular that the quantum instability 
for concentric fuzzy spheres disappears for $D=10$.
%
This is interesting since
concentric fuzzy spheres with different radii 
are identified with the 't Hooft-Polyakov monopoles \cite{0412052}
(See also ref.\ \cite{Aoki:2006wv}.).
Such configurations are perturbatively stable
in the bosonic model, but not always in the supersymmetric models.



{}From the string theoretical viewpoint \cite{alekseev}, 
the classical instability corresponds to the appearance 
of tachyons in the spectrum of an open string 
connecting the intersecting fuzzy spheres. 
The process of the tachyon condensation can be studied
by Monte Carlo simulation as in ref.\ \cite{0401038},
but we do not pursue it here.
The quantum instability for large separation,
on the other hand, 
is due to the attractive force induced by the closed string
propagation between the fuzzy spheres.

The rest of this article is organized as follows. 
In section \ref{fuzzymodel}
we discuss the stability of 
the multi-fuzzy-sphere configurations
in the simplest model.
In sections \ref{susy_fuzzy} and \ref{higher_fuzzy}
we extend the analysis to 
the supersymmetric and higher-dimensional models. 
Section \ref{summary} is devoted to a summary 
and discussions.

\section{The $D=3$ bosonic model
}\label{fuzzymodel}

In this section we investigate
the properties of multi-fuzzy-sphere configurations
in the simplest dimensionally reduced YMCS model.
We calculate the one-loop effective action 
around those configurations,
and discuss their stability.

\subsection{The model and its classical solutions}\label{classicalpic}
\label{model-solutions}

We consider the model 
defined by the action \cite{alekseev,0101102}
\bel{b-ac}
 S= N \, \tr \Big(
-\frac{1}{4} \, [ A_{\mu} ,A_{\nu}] [A_{\mu} ,A_{\nu}]
  + \frac{2}{3} \, i \, \alpha \, \epsilon_{\mu \nu \lambda}
A_{\mu} A_{\nu} A_{\lambda}  \Big) \ ,
\eeq
which can be obtained by taking the zero-volume limit
of the 3d YMCS
theory.
The $N\times N$ matrices $A_\mu$ are traceless Hermitian, 
and the Greek indices run over 1 through 3.
The existence of the Chern-Simons term makes it possible
for the model to have various types of
fuzzy-sphere configurations as classical solutions.

The action (\ref{b-ac}) has the SO$(3)$ symmetry,   
the ``translational symmetry'' 
$ A_{\mu} \rightarrow  A_{\mu}+ \alpha_\mu {\bf 1}$ 
and the SU($N$) symmetry 
$A_{\mu} \rightarrow UA_{\mu}U^{\dagger}$.
In ref.\ \cite{0310170} it is shown that the convergence 
property of the path integral over the non-compact 
dynamical variables
$A_\mu$ \cite{Krauth:1998yu,9811220,Austing:2001bd}
is not affected by the addition 
of the Chern-Simons term. Therefore, the path
integral of this model converges for $N \geq 4$.

The classical equation of motion is obtained as
\beqa
\Bigl[A_\nu, [ A_\nu, A_\mu ] \Bigr]
 + i \, \alpha \, \epsilon_{\mu\nu\lambda}
[A_\nu, A_\lambda ] = 0 \ . 
\label{eom1}
\eeqa
The 
general 
solution takes the form
\beq
 A_\mu = X_\mu \equiv  \alpha \, \bigoplus_{I=1}^{k} 
\Bigl(     L_\mu^{(n_I)} +  
x_\mu^{(I)} \, {\bf 1}_{n_I}
\Bigr) \ ,
\label{general-sol}
\eeq
where $L^{(n)}_\mu$ is the representation matrix
for the $n$-dimensional irreducible representation of the SU$(2)$ 
algebra
$[L^{(n)}_\mu,L^{(n)}_\nu ]   =  i \, 
\e_{\mu \nu \l} L^{(n)}_\l $,
and $\sum_{I=1}^k n_I = N$.
Due to the identity
\bel{fuzeqn2}
\sum_{\l=1}^3 \left(
L_\l^{(n)}  \right)^2 = \frac{1}{4} \, 
(n^2-1) \, {\bf 1}_{n} \ ,
\eeq
we may consider the configuration (\ref{general-sol}) 
as representing $k$ fuzzy spheres with 
the radii 
\beq
r_{I}= \frac{1}{2} \sqrt{(n_I)^2-1}
\label{def-radii}
\eeq
and the center at $x_\mu ^{(I)}$.
Here and henceforth, we measure the length in units of $\alpha$.
Plugging this solution into (\ref{b-ac}),
we obtain the classical part of the effective action
\begin{eqnarray}
W_{\rm cl} = 
- \frac{\alpha^{4} N}{24} \sum_{I=1}^{k} 
\Bigl \{(n_{I})^{2} - 1 \Bigr\} \ .
\label{eff-cl}
\end{eqnarray}

\subsection{One-loop effective action}
\label{oneloops}

Next we calculate the effective action
around the classical solutions.
At the one-loop level,
it suffices to consider the interaction 
between two fuzzy spheres.
The result for the multi-fuzzy-sphere
configuration (\ref{general-sol}) can be
readily obtained by summing over all possible
pairs of fuzzy spheres.
Therefore, we restrict ourselves in what follows 
to the two-fuzzy-sphere 
configuration; i.e., 
the $k=2$ case in eq.\ (\ref{general-sol}) given by
\beq
 X_\mu = \alpha \left(
\begin{array}{cc}
 L_\mu^{(n_1)} + x_\mu ^{(1)} {\bf 1}_{n_1}   &  ~ \\
 ~  &   L_\mu^{(n_2)} + x_\mu ^{(2)} {\bf 1}_{n_2}
\end{array}
\right)  \ .
\label{two-fuzz-bg}
\eeq
Due to the
``translational symmetry'' mentioned 
in section \ref{model-solutions},
the result will depend only on the displacement vector 
\beq
\xi_\mu =
x_\mu^{(1)} - x_\mu^{(2)} \ .
\eeq
Furthermore, exploiting the SO(3) symmetry, 
we may restrict ourselves
to the case $\xi_{\mu} = (0,0,\xi)$
without loss of generality.
We will therefore obtain the one-loop effective action
as a function of a single parameter $\xi$.
%

We expand the original matrices around the background 
(\ref{two-fuzz-bg}) as
\beq
A_\mu=X_\mu + \tilde{A}_\mu \ .
\label{def-fluct}
\eeq
We add the gauge fixing term and the ghost term
\begin{eqnarray}
S_{\rm g.f.} = -\f{N}{2} \tr \, [X_\mu, A_\mu]^2, ~~~~
S_{\rm gh} = - N  \tr \, [X_\mu, {\bar c}][A_\mu, c] \ . 
\label{gf}
\end{eqnarray}
Plugging (\ref{def-fluct}) into the actions
(\ref{b-ac}) and (\ref{gf}), we obtain the quadratic terms
\beqa
S_2 &=&\frac{1}{2}\,  N \, \tr \left( 
\tilde A_\mu [X_\lambda , [X_\lambda , \tilde A_\mu ]] \right)
+ N \, \tr \Bigl( \bar c \, [X_\lambda , [X_\lambda , c ]] \Bigr)
\nonumber\\&& 
-\, N \, \tr \Bigl\{
\Bigl( [X_\mu , X_\nu] - i \alpha \epsilon_{\mu\nu\rho} X_\rho \Bigr)
[\tilde A_\mu , \tilde A_\nu ]\Bigr\} \ .
\label{quad-terms}
\eeqa
Corresponding to the two-fuzzy-sphere 
configuration (\ref{two-fuzz-bg}),
we decompose the fluctuation matrices as
\begin{eqnarray}
& & \tilde{A}_\mu=\left(
\begin{array}{cc}
 a_\mu^{(1)}   & b_\mu\\
b_\mu^\dag & a_\mu^{(2)}
\end{array}
\right) \ , 
\quad
{\bar c}=\left(
\begin{array}{cc}
{\bar c}^{(1)} &  \beta \\
\bar{\gamma} & {\bar c}^{(2)}
\end{array}
\right) \ ,
\quad
c=\left(
\begin{array}{cc}
c^{(1)} & \mbox{~}\gamma \\
- \bar{\beta} & \mbox{~}c^{(2)}
\end{array}
\right) \ , 
\label{fluct}
\end{eqnarray}
where the first and second diagonal blocks are
$n_1 \times n_1$ and $n_2 \times n_2$ matrices, respectively.
Plugging (\ref{two-fuzz-bg}) and (\ref{fluct}) into 
(\ref{quad-terms}), we obtain
\beqa
S_2 &=& S_{\rm 2}^{\rm (self)} +
S_{\rm 2}^{\rm (int)} \ ,
 \nonumber \\
S_{\rm 2}^{\rm (self)} &=& N \a^2 \sum_{I=1,2}
\left[ -\frac{1}{2}
   \tr \left(
 [ L^{(n_I)}_\mu , a^{(I)}_\nu  ]^2
\right) +    \tr
\Bigl\{ [L^{(n_I)}_\mu, {\bar c}^{(I)}]
[L^{(n_I)}_\nu, c^{(I)}] \Bigr\} \right]
 \ , \label{eqng} \\
S_{\rm 2}^{\rm (int)} &=& 
N \a^2 \Bigl\{
b_\mu^\dag
\left( \mathcal{H}^2 \delta_{\mu\nu}
-2i \e_{\mu\nu\l}
\xi_\l 
\right)
b_\nu + \bar{\beta}
\mathcal{H}^2
\beta
+ \bar{\gamma}
\mathcal{H}^2
\gamma  \Bigr\} \ , 
\label{intg}
\eeqa
where we have introduced
a linear operator
\beqa
\mathcal{H}_\mu &=& \mathcal{J}_\mu + \xi_\mu   \ , \\
\mathcal{J}_\mu&=& 
L_\mu^{(n_1)} \otimes {\bf 1}_{n_2} + 
{\bf 1}_{n_1} \otimes (- L_\mu^{(n_2)*}) \ ,
\eeqa
which act on the space of $n_1 \times n_2$ matrices.
%

The one-loop effective action $W$ is defined by
\beq
\ee^{-W}= \int da\, db\,  dc\, d {\bar c}\, d\beta \,
d \bar{\beta}  \, d\gamma\, d \bar{\gamma} \, \ee^{-S_2} \ .
\eeq
In particular, the terms that come from
the interaction part of the action are given by
\beq
W_{\rm int}
= \log \det 
\frac{(\mathcal{H}^2+2 \xi)(\mathcal{H}^2-2 \xi)}
{\mathcal{H}^2} 
\ , 
\label{c-1loop}
\eeq
where we have omitted a $\xi$-independent term.

Thus the calculation reduces to the eigenvalue problem
of the $\mathcal{H}^2$, which is given by
\beq
\mathcal{H}^2 = \mathcal{J}^2 + 2 \, \xi  \mathcal{J}_3 + \xi^2  \ .
\eeq
This can be readily solved \cite{0108002}
by noticing that $\mathcal{J}_\mu$ can be regarded
as the total angular momentum operator of the system
composed of spins $j_1=\frac{n_1-1}{2}$ and $j_2=\frac{n_2-1}{2}$.
Hence $\mathcal{J}^2$ and $\mathcal{J}_3$ are
simultaneously diagonalizable and 
their eigenvalues are given by $j(j+1)$ and $m$, respectively,
where $j$ and $m$ take
\beqa
j &=& j_{\rm min} , j_{\rm min}+1 , \cdots  , j_{\rm max}  \ , \\
m &=&  -j , -j+1 , \cdots , j  
\eeqa
with $j_{\rm min}=|j_1-j_2|=\frac{|n_1-n_2|}{2}$ and 
$j_{\rm max}=j_1+j_2 = \frac{n_1+n_2}{2}-1$.
Therefore, the eigenvalues of the operator 
$\mathcal{H}^2$ are given by
\beq
h(j,m) =  \xi^2 + 2 \, \xi \, m + j \, (j+1)  \ ,
\label{h-def}
\eeq
and the effective action (\ref{c-1loop})
is obtained as
\beqa
\label{Wint-expr}
W_{\rm int} 
&=& 
\sum_{j=j_{\rm min}}^{j_{\rm max}}  \log w_j  \ ,  \\
w_j &=&  
\prod_{m=-j}^{j} \frac{h(j,m+1) \, h(j,m-1)}{h(j,m)}
= \frac{h(j,j+1) \, h(j,-j-1)}{h(j,j) \, h(j,-j)} 
\prod_{m=-j}^{j} h(j,m) \ .
%
%
\label{Wresult}
\eeqa
In the case of concentric fuzzy spheres ($\xi=0$), we get
\begin{equation}
W_{\rm int}
= \sum_{j=j_{\rm min}}^{j_{\rm max}} (2 j +1) 
\log \left[ j(j+1) \right]  \ ,
\label{Wintlog}
\end{equation}
which agrees with the result of 
refs.\ \cite{0412052,dyn-gauge}.

If we further consider the case
with equal radii ($n_1=n_2=n$),
which corresponds to coinciding fuzzy spheres,
we have $j_{\rm min}=0$ and $j_{\rm max}= n-1$. 
Note that the argument of the $\log$ in (\ref{Wintlog})
vanishes for $j=0$, 
which indicates the appearance of zero modes.
The fate of these zero modes is 
discussed in refs.\ \cite{0401038,0412052}
in detail.
In what follows, we therefore exclude
this case.
Then all the eigenvalues $h(j,m)$ of the operator $\mathcal{H}^2$
are strictly positive.

\subsection{Stability of the 
classical solutions
} 
\label{metastable_fuzzy}


First we discuss the classical stability of 
the two-fuzzy-sphere configuration.
Instability can appear only from the operator $\mathcal{H}^2-2\xi$.
Its eigenvalue $h(j,m)-2\xi$ is negative if and only if
$m=-j$ and $\xi_{j,-} < \xi < \xi_{j,+}$, where
\begin{eqnarray}
\xi_{j,\pm} = (j+1) \pm \sqrt{j+1}  \ .
\label{xipm-def}
\end{eqnarray}
Since $\xi_{j,\pm}$ increases monotonically
with $j$ and $\xi_{j+1,-} < \xi_{j,+}$, we conclude that
the two-fuzzy-sphere configuration is unstable for
\begin{eqnarray}
\xi_{j_{\rm min},-} < \xi < \xi_{j_{\rm max},+}  \ . 
\label{inst-region}
\end{eqnarray}
Note that the both ends of the region is given approximately by
\beq
\xi_{j_{\rm min},-} \approx 
|r_1 - r_2 |   \ ,   \quad \quad
\xi_{j_{\rm max},+} \approx 
r_1 + r_2 
\eeq
for large $n_1$ and $n_2$, which also implies 
large $N$($=n_1+n_2$).
Therefore, eq.\ (\ref{inst-region}) implies that
intersecting fuzzy spheres are unstable.
At finite $N$, however, 
the second term in eq.\ (\ref{xipm-def}) is
non-negligible, and we have instability 
even when the two fuzzy spheres are
close to intersecting. 

Next let us discuss the quantum stability
by looking at the $\xi$-dependence of the one-loop effective
action in the region outside (\ref{inst-region}).
We take the $N \rightarrow \infty$ 
limit in such a way that the radii $r_1$, $r_2$ 
given by (\ref{def-radii})
and the distance $\xi$ are of the same order.
For that purpose, it is convenient to introduce 
the parameters
\beq
\nu \equiv \frac{|n_{1} - n_{2}|}{N} \ ,
\quad \quad
\tilde{\xi} \equiv \frac{\xi}{N} \ ,
\label{nu-def}
\eeq
which corresponds to
\beq
\nu \approx \frac{|r_1 - r_2|}{r_1 + r_2} \ ,
\quad \quad
\tilde{\xi} \approx \frac{\xi}{2 \, (r_1 + r_2)} \ .
\label{nu-def2}
\eeq
We therefore take the large-$N$ limit fixing 
$\nu$ and $\tilde{\xi}$.
In that limit the sum over $m$,
which appears after taking the log of (\ref{Wresult}),
and the sum over $j$ in (\ref{Wint-expr})
can be replaced by integrals, and we obtain
\beqa
W_{\rm int} 
& \simeq & N^2 \left\{ 
F\left(\tilde{\xi} , \frac{1}{2}\right) 
- F\left(\tilde{\xi} , \frac{\nu}{2} \right) 
\right\}  \ , \\
F(\tilde{\xi} , x) 
&=& ( 2 \log N - 1 ) \, x^2
+ \frac{1}{\tilde{\xi}}
\left\{ f\left(x +\tilde{\xi}\right)
- f\left(x -\tilde{\xi} \right) \right\}  \ , \\
f(x) &=& \frac{1}{6} x^3 \log x^2 - \frac{1}{9} x^3 \ .
\eeqa
Note that this expression
is not valid for 
$\frac{\nu}{2} < \tilde{\xi} < \frac{1}{2}$,
which corresponds to the region
of classical instability (\ref{inst-region}). 
Outside that region, $W_{\rm int}$ is a monotonically increasing 
function of $\tilde{\xi}$ for any $0 < \nu < 1$.
The asymptotic behavior of $W_{\rm int}$ is obtained as
\beq
W_{\rm int} \simeq
\left\{ 
\begin{array}{ll}
\frac{2}{3} N^2 \log (\nu^{-1}) \, 
\tilde{\xi}^2  & \mbox{~for~}\tilde{\xi} \ll \frac{\nu}{2}  
\ , \\
\frac{1}{2}  N^2 (1 - \nu ^2 )  \log \tilde{\xi}
 & \mbox{~for~}\tilde{\xi} \gg \frac{1}{2}  \ ,
\end{array}
\right. 
\label{asym-boson}
\eeq
where we have omitted irrelevant constant terms.

When the two fuzzy spheres are located
away from each other,
they attract each other until they touch
and run into the classical instability.
When one fuzzy sphere is inside the other,
they tend to become concentric.
Therefore,
concentric fuzzy spheres
with different radii,
which are identified with the 't Hooft-Polyakov
monopoles \cite{0412052},
are perturbatively stable in the bosonic model.

\section{Supersymmetric model} 
\label{susy_fuzzy}

In this section we extend our analysis
in the previous section
to a supersymmetric version of the model (\ref{b-ac}),
which is defined by \cite{alekseev,0101102}
\bel{fuzaction}
 S= N \, \tr \left(
-\frac{1}{4} \, [ A_{\mu} ,A_{\nu}] [A_{\mu} ,A_{\nu}]
  +\frac{2}{3} \, i \,  \alpha \, \epsilon_{\mu \nu \lambda}
A_{\mu} A_{\nu} A_{\lambda}
  +\frac{1}{2} \psi_{\alpha} (\Gamma_{\mu})_{\alpha\beta}
[A_{\mu},\psi_\beta]  \right) \ ,
\eeq
where 
$\psi_\alpha$ ($\alpha = 1 , 2$)
is a two-component Majorana spinor, 
each component being a 
$N \times N$ traceless Hermitian matrix.
The $2 \times 2$ matrix
$\Gamma_\mu= \mathcal{C} \gamma_\mu$ ($\mu=1,2,3$)
is a product of 
the charge conjugation matrix $\mathcal{C}$
and the Euclidean gamma matrix $\gamma_\mu$.
This action has a ${\cal N}=2$ supersymmetry \cite{0101102}.
For $\alpha=0$, the path integral is known to be divergent 
\cite{Krauth:1998xh,Austing:2001pk}, which
also applies to the $\alpha \neq 0$
case \cite{0310170}. 
This problem does not occur
in higher dimensional models, which we discuss in the
next section.
However, even for $D=3$, we can still 
make a well-defined perturbative expansion
around the general solution (\ref{general-sol}).
As in the bosonic case, it suffices to consider the 
two-fuzzy-sphere configuration (\ref{two-fuzz-bg})
at the one-loop level.
The region of classical instability (\ref{inst-region})
corresponding to intersecting fuzzy spheres remains the same,
but we will see below that 
the one-loop effective action, and hence the issue 
of quantum stability changes drastically
due to the existence of supersymmetry.

We decompose the fermionic matrix $\psi_{\alpha}$ as
 \begin{eqnarray}
   \psi_{\alpha} = \left( \begin{array}{cc} s^{(1)}_{\alpha}
    & t_{\alpha} \\ t^{\dagger}_{\alpha} & s^{(2)}_{\alpha} 
\end{array} \right) \ .
 \end{eqnarray}
The contribution of the fermions to the quadratic action is 
\beqa
S_{\rm 2,F}^{\rm (self)}&=& \frac{1}{2} \a N
(\Gamma_\mu)_{\alpha\beta}
\sum_{I=1,2} \tr \left\{s_\alpha^{(I)}  
[L^{(I)}_\mu, s_\beta^{(I)}] \right\} \ , 
\label{skin-f} \\
S_{\rm 2,F}^{\rm (int)}
&=& 
\a N (\Gamma_\mu)_{\alpha\beta}
\, \tr \, 
(t_\alpha^\dagger \, \mathcal{H}_\mu t_\beta ) \ . \label{ikin-f}
\eeqa
{}From the integration over the matrix $t_\alpha$,
we get $\det(\Gamma_\mu \mathcal{H}_\mu )$.
This gives an extra term
\beq
W_{\rm int}^{\rm (F)} =
- \log | \det(\Gamma_\mu \mathcal{H}_\mu )|
= - \log | \det (\sigma_\mu \mathcal{H}_\mu ) | 
\label{c-1loop-f} 
\eeq
to the effective action (\ref{c-1loop}),
where $\sigma_\mu$ are the Pauli matrices.

The evaluation of the determinant is slightly more
involved than in the bosonic case.
For that we add a spin $\frac{1}{2}$ system
to the previously considered spin $j$ system.
The total angular momentum operator is given by
\beq
\mathcal{K}_\mu
= \mathcal{J}_\mu + \frac{\sigma_\mu}{2} \ .
\eeq
As the basis of the combined system,
we use the eigenstates $|k,n \rangle$
of the operators
$\mathcal{K}^2$ and $\mathcal{K}_3$
with the eigenvalues $k(k+1)$ and $n$, respectively,
where $k=j \pm \frac{1}{2}$ and $n=-k ,\cdots , k$.
We note that
\beq
\sigma_\mu \mathcal{H}_\mu
= \mathcal{K}^2 - j(j+1) - \frac{3}{4} + \xi \, \sigma_3 \ .
\eeq
Due to the last term, the operator is not
diagonalized with the chosen basis.
However, since $\sigma_3$ commutes with $\mathcal{K}_3$,
the last term only mixes the states with the same $n$.
This means that 
for $|n| \le j-\frac{1}{2}$, 
only the two states $|j+\frac{1}{2} ,n \rangle$ and 
$|j-\frac{1}{2} ,n \rangle$ are mixed.
Using the Clebsch-Gordan coefficients, 
the corresponding matrix elements are obtained as
 \beqa
\Big\langle j+\frac{1}{2}, n \Big| \sigma_\mu \mathcal{H}_\mu 
\Big| j+\frac{1}{2} ,n \Big\rangle  
&=& j+ \frac{2n\xi}{2j+1}  \ , \\
\Big\langle j\pm\frac{1}{2}, n \Big| \sigma_\mu \mathcal{H}_\mu 
\Big| j\mp\frac{1}{2} ,n \Big\rangle  
&=& -  \frac{2\xi}{2j+1} 
\sqrt{\left(j-n+\frac{1}{2}\right)\left(j+n+\frac{1}{2}\right)}
\ ,  \\ 
\Big\langle j-\frac{1}{2}, n \Big| \sigma_\mu \mathcal{H}_\mu 
\Big| j-\frac{1}{2} ,n \Big\rangle
&=& -j - 1 - \frac{2n\xi}{2j+1}  \ .
\eeqa
The determinant of the $2 \times 2$ matrix is given
by $h(j,n)$ using the notation (\ref{h-def}).
For $|n| = j+\frac{1}{2}$, there is no mixing
and we get
\beq
\Big\langle j+\frac{1}{2}, \pm \left(j+\frac{1}{2}\right)
 \Big| \sigma_\mu \mathcal{H}_\mu 
\Big| j+\frac{1}{2}, \pm \left(j+\frac{1}{2}\right) \Big\rangle  
= j \pm \xi \ .
\eeq
Therefore, we obtain
\beq
 W_{\rm int}^{\rm (F)} =  
- \sum_{j=j_{\rm min}}^{j_{\rm max}} \log 
 w_j^{\rm (F)}  \ , \quad \quad
w_j^{\rm (F)}
=  |\xi^2 -j^2 | \prod_{n=-j+\frac{1}{2}}^{j-\frac{1}{2}} 
h(j,n)  \ .
\label{wf}
\eeq
Adding this to the previous result, we obtain
the total effective action (\ref{Wint-expr}), 
where $w_j$ is now replaced by
\beqa
w_j
&=& 
\frac{h(j,j+1) \, h(j,-j-1)}{h(j,j) \, h(j,-j)} 
\times  \left\{ 
\frac{h(j,j+\frac{1}{2})}{|\xi ^2 - j^2|}
\prod_{m=-j}^{j} \frac{h(j,m)}{h(j,m+\frac{1}{2})} 
\right\}  \ .
\label{W-higherD}
\eeqa

In the large-$N$ limit,
we find that the O($N^2$) term and the O($N$) term 
vanish exactly due to supersymmetry, and we are left
with an O(1) quantity.
Its asymptotic behavior can be obtained as 
\beq
W_{\rm int}^{\rm (SUSY)} \simeq
\left\{ 
\begin{array}{ll}
\log (\nu^{-1})  -  12 (\nu^{-2} -1) 
\tilde{\xi}^{2} + {\rm O}(\tilde{\xi}^4)  
  & \mbox{~for~}\tilde{\xi} \ll \frac{\nu}{2}    
\ , \\
- (1 - \nu^2) \tilde{\xi}^{-2} 
+ {\rm O} ( \tilde{\xi}^{-4}) 
 & \mbox{~for~}\tilde{\xi} \gg \frac{1}{2}    \ ,
\end{array}
\right. 
\label{asym-susy}
\eeq
which should be compared with the results
(\ref{asym-boson}) for the bosonic case.
The factor of $N^2$ is absent in (\ref{asym-susy}),
and therefore
the interaction is much weaker.
The interaction at small $\tilde{\xi}$
is repulsive, which implies that 
the concentric fuzzy sphere has quantum instability
in contrast to the bosonic case.

\section{Higher-dimensional models} 
\label{higher_fuzzy}

Let us further extend our analysis to
higher dimensional models defined by the action
\bel{fuzactionDdim}
 S= N \, \tr\Big(
-\frac{1}{4} \, [ A_{\mu} ,A_{\nu}] [A_{\mu} ,A_{\nu}]
  +\frac{2}{3} \, i \, \alpha \, \epsilon_{a b c}
A_{a}A_{b}A_{c}
 +\frac{1}{2} \, \bar{\psi} \, 
\Gamma_{\mu}[A_{\mu},\psi]  \Big) \ ,
\eeq
where the indices run over $\mu,\nu=1,2,\cdots,D$ and $a,b,c=1,2,3$.
The dimensionality is limited to $D=3,4,6,10$, 
where the matrix model (\ref{fuzactionDdim}) has supersymmetry
with an appropriate choice of the spinor representations for the
fermions.
The $D=4$ case has been studied by Monte Carlo
simulation in ref.\ \cite{0506062}.

As a classical solution,
we take the configuration (\ref{two-fuzz-bg}) for $\mu=1,2,3$
and $X_\mu = 0$ otherwise.\footnote{In the higher dimensional
models, one can separate the two fuzzy spheres
also in the $4,5, \cdots , D$ directions.
We do not discuss the results here, since they
are not very illuminating.
}
This describes a system of two fuzzy spheres 
extended in the $1,2,3$ directions
of the $D$-dimensional target space. 
The region of classical instability is the same as 
(\ref{inst-region}).
Outside that region,
the interaction part of the one-loop effective action
is given as
\beq
W_{\rm int}^{{\rm (SUSY)},D}
= \log  \det 
\left(\mathcal{H}^2-2\xi\right)
\left(\mathcal{H}^2+2\xi\right) 
(\mathcal{H}^2)^{D-4} 
- \log \left| \det (\sigma_\a \mathcal{H}_a)^{D-2} \right| \ .
\eeq
This expression can be evaluated
as in the previous section,
and the generalization simply amounts to modifying 
(\ref{W-higherD}) by raising the second factor 
in the parenthesis $\{ \ \} $ to the power of $(D-2)$.
Thus we obtain
\beq
W_{\rm int}^{{\rm (SUSY)},D} \simeq
\left\{ 
\begin{array}{ll}
(D-2) \log (\nu^{-1})  + 4 \, (D-6) (\nu^{-2} -1) 
\tilde{\xi}^{2} + {\rm O}(\tilde{\xi}^4)  
  & \mbox{~for~}\tilde{\xi} \ll \frac{\nu}{2}  
\ , \\
- (1 - \nu^2) \tilde{\xi}^{-2} 
+ {\rm O} ( \tilde{\xi}^{-4}) 
 & \mbox{~for~}\tilde{\xi} \gg \frac{1}{2}  \ .
\end{array}
\right. 
\label{asym-susyD}
\eeq
It is interesting that the coefficient of the 
$\tilde{\xi}^{2}$ changes its sign at $D=6$.
Therefore, concentric fuzzy spheres are
perturbatively stable 
in the $D=10$ supersymmetric model.

\section{Summary and discussions} 
\label{summary}
In this paper we have discussed the classical and
quantum stability of 
the multi-fuzzy-sphere configurations in the YMCS models 
using the one-loop effective action. 
We have shown in general that 
the 
configuration becomes
unstable when fuzzy spheres intersect.
Separate fuzzy spheres attract each other in general,
and eventually run into instability upon intersecting.
On the other hand, the fate of configurations with
one fuzzy sphere located inside another depends on the model.
We find that the concentric fuzzy spheres,
which correspond to the 't Hooft-Polyakov monopoles,
are perturbatively stable in the bosonic model and 
in the $D=10$ supersymmetric model.

Our ambitious goal is to investigate
the dynamical generation of gauge group,
as well as that of the space-time, 
in nonperturbative formulations of superstring
theory such as the IIB matrix model. 
Refs.\ \cite{cosetspace} present a closely related
approach, in which one attempts to obtain the 
standard model gauge group using fuzzy spheres
in the extra dimensions. There, however, the
4d space-time is introduced from the outset 
as in ordinary field theories.
We hope that the dynamical properties of
fuzzy spheres studied in this paper will be
useful in understanding how our universe has emerged.

\acknowledgments
We would like to thank K. Nagao for helpful discussions. 



\end{document}